\documentclass[twocolumn,secnumarabic,amssymb, nobibnotes, aps, prd, superscriptaddress]{revtex4-1}

\usepackage{graphicx}
\usepackage{amsmath}
\usepackage{upgreek}
\usepackage[monochrome]{color}

\DeclareTextSymbol{\degres}{OT1}{23}

\begin{document}
\title{Friction of polymers: from PDMS melts to PDMS elastomers}
\author{Marceau H\'enot}
\affiliation{Laboratoire de Physique des Solides, CNRS, Univ. Paris-Sud, Universit\'e
Paris-Saclay, 91405 Orsay Cedex, France}
\author{\'Eric Drockenmuller}
\affiliation{Univ Lyon, Universit\'e Lyon 1, CNRS, Ing\'enierie des Mat\'eriaux Polym\`eres, UMR 5223, F-69003, Lyon, France}
\author{Liliane L\'eger}
\affiliation{Laboratoire de Physique des Solides, CNRS, Univ. Paris-Sud, Universit\'e
Paris-Saclay, 91405 Orsay Cedex, France}
\author{Fr\'ed\'eric Restagno}
\email[Corresponding author : ]{frederic.restagno@u-psud.fr}
\affiliation{Laboratoire de Physique des Solides, CNRS, Univ. Paris-Sud, Universit\'e
Paris-Saclay, 91405 Orsay Cedex, France}
\date{\today}
\begin{abstract}
The slip behavior of polydimethylsiloxane (PDMS) polymer melts flowing on weakly adsorbing surfaces made of short non-entangled PDMS chains densely end-grafted to silica has been characterized. For high enough shear rates, slip lengths proportional to the bulk fluid viscosity have been observed, in agreement with Navier's interfacial equation, and demonstrating that the interfacial Navier's friction coefficient is a local quantity, independent of the polymer molecular weight. Comparing the interfacial shear stresses deduced from these measured slip lengths to available friction stress measured for crosslinked PDMS elastomers, we further demonstrate the local character of the friction coefficient and compare its value to the monomer-monomer friction. 
\end{abstract}
\maketitle

Understanding the mechanical properties of polymer melts or polymer solutions usually needs a rheometer in which the friction on the plates is supposed to be known in order to decouple the interfacial properties and the bulk properties of the polymer. The simplest hypothesis is to assume the relative tangential fluid velocity to be zero. Historically, Navier introduced in a linear response approach a more general condition \cite{navier_memoire_1823}: the shear stress at the solid-liquid interface should be proportional to the component of the fluid velocity tangent to the surface $V$:
\begin{equation}
\sigma_\mathrm{fluid\rightarrow surface}=k V
\label{eq_contrainte_bord}
\end{equation}
where $k$ is an interfacial solid-liquid friction coefficient, assumed to be independent of the shear rate. This coefficient is usually converted into the so called Navier's slip length $b=\eta/k$. It represents the distance from the surface at which the velocity profile extrapolates to zero. The determination of slip lengths for simple fluids has been the subject of intensive experimental~\cite{chan_drainage_1985,pit_direct_2000,cottin-bizonne_boundary_2005,schmatko_friction_2005,neto_boundary_2005,lauga_2007} and theoretical \cite{bocquet_hydrodynamic_1994,thompson_general_1997,priezjev_molecular_2004, bocquet_flow_2007} researches over the last 20 years. 
The measured slip lengths lie between 0 to 50~nm and appear to be highly sensitive to tiny molecular details of the surfaces~\cite{pit_direct_2000,schmatko_friction_2005}.
	
Polymer melts can present huge slip lengths contrary to simple liquids. They thus are interesting candidates to quantitatively test Eq.~\ref{eq_contrainte_bord} and possibly get rid of the molecular details of the surface. Indirect experimental evidences for a giant slip of polymers melts were provided by extrusion instabilities reported since the 40's~\cite{petrie_instabilities_1976}, extensively studied by polymer rheologists~\cite{el_kissi_different_1990}. In 1979, de Gennes proposed a simple explanation of such a huge slip, for polymer melts~\cite{de_gennes_1979} flowing on ideal non-adsorbing surfaces. The physical idea is that $k$, which results from the local contact between monomers and the solid wall, should be independent of chain entanglements and chain length, while entanglements do control the polymer melt viscosity. The slip length should thus simply be proportional to the polymer viscosity:
\begin{equation}
b(M)=\frac{\eta(M)}{k}
\label{eq_slip_length}
\end{equation}
where $M$ is the molecular weight of the polymer. 
Few attempts to test quantitatively Eq.~\ref{eq_slip_length} have been reported despite the potential interest in terms of testing the Navier's hypothesis. An important experimental issue is the strong tendency of polymer molecules to adsorb at surfaces. Adsorption of few polymer molecules on the surface has been reported to be responsible for a strongly non linear slip behavior, characterized by three different slip regimes depending on the shear rate: for low shear rates, entanglements between surface adsorbed and bulk chains produce a large friction and thus a weak slip, while for high enough shear rates, bulk and surface chains become stretched and fully disentangled, leading to high slip. In between, the slip length increases at characteristic fixed shear rate~\cite{leger_wall_1997,mhetar_slip_1998,massey_1998,fetzer_slip-controlled_2006,brochard_1992,migler_slip_1993,brochard-wyart_slippage_1996}. 
Durliat~\textit{et al.}~\cite{durliat_influence_1997} observed a shear independent high slip regime, for one PDMS molecular weight flowing on weakly dense PDMS grafted layers. Mhetar and Archer \cite{mhetar_slip_1998} investigated the molecular weight dependence and measured  slip lengths scaling like $b\sim M^{1.3}$ on weakly adsorbing surfaces due to some defects on their surfaces. Wang and Drda \cite{wang_molecular_1997} reported a slip length at the onset of high slip scaling like $b \sim M^{3.5}$. More recently, Dewetting experiments \cite{baumchen_reduced_2009,baumchen_slippage_2012}, reported slip lengths proportional to the viscosity for high molecular weights melts. All these studies indicate that high slip can indeed be obtained with polymer melts at large enough shear rates and tend to show that the slip lengths do scale with bulk viscosities. They however lack two crucial facets needed to understand the phenomenon at the molecular level: a clear investigation of the shear rate dependence, and an independent determination of the interfacial friction coefficient, $k$, so that the validity of the Navier equation could be fully tested. 

We present here direct proofs of both Navier and de Gennes hypothesis, based first on direct measurements of the slip lengths for PDMS melts with various molecular weights and narrow molecular weight distributions, flowing on silica surfaces decorated with densely packed end-anchored non-entangled short PDMS chains, and second, on direct friction measurements of crosslinked PDMS elastomers sliding on the same surfaces.

All polymer fluids used were trimethylsiloxy terminated PDMS melts with number average molecular weights $M_{\mathrm n}$ of 610, 787 and 962~kg\(\cdot\)mol$^{-1}$ with polydispersity indexes 1.15, 1.22 and 1.27 respectively. These melts were obtained by controlled fractionation of a commercial batch (Petrarch PS047.5), and mixed with 0.5$\%$ by weight of fluorescently labled photobleachable PDMS chains with a number average molecular weight $M_{\mathrm n}^\star = 321$~kg\(\cdot\)mol$^{-1}$ and polydispersity index 1.18. \textcolor{red}{At this low concentration, the labeled chains aee known not to affect the rheology of the system~\cite{rhodorsil}.} fluorescent chains were lab-synthesized and labeled at both ends with nitrobenzoxadiazole groups (NBD) emitting at 550~nm~\cite{leger1996,cohen_synthesis_2012} when excited at 458~nm. The surfaces on which slip was investigated were the polished surface of a fused silica prism, covered with end grafted short PDMS chains with an average number molecular weight $2\times 10^{3}$~g$\cdot$mol$^{-1}$, well below the average molecular weight between entanglements, $M_\mathrm{e}\approx 10\times 10^{3}$~g$\cdot$mol$^{-1}$ for PDMS ~\cite{fetters_1994,Leger1999}. 

The synthesis protocol of these end-functionalized chains along with the grafting procedure are detailed in Marzolin~\textit{et al.}~\cite{marzolin_2001}. The dry thickness of the grafted layer was 3.2~nm which corresponds to grafted chains in the stretched regime. 
The advancing contact angle of water of this surface was $\theta_\mathrm{a}=112^\circ$ with an hysteresis of $5^\circ$. 

\begin{figure}[htbp]
  \centering
  \includegraphics[width=8cm]{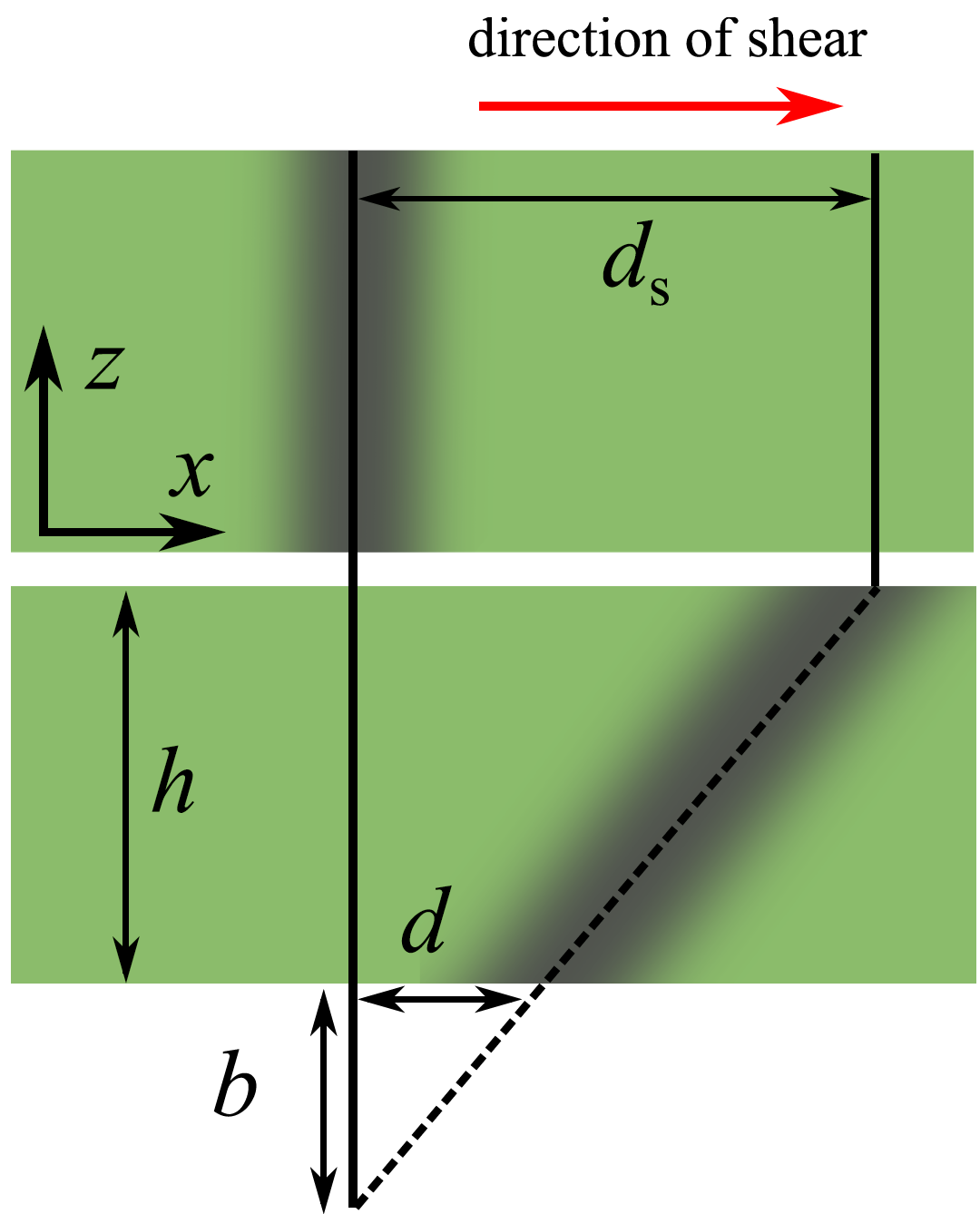}
  \caption{Principle of velocimetry using photobleaching. Top: schematics of the photobleached line inside in the fluid. Bottom: the liquid has been sheared over a distance $d_\mathrm{s}$, with a no slip boundary condition at the top plate while it has slipped at the bottom plate over a distance $d$ which corresponds to a slip length $b$. The observation of the photobleached line from the top allows to measure $b$ and the shear rate $\dot{\gamma}$.}
  \label{principe_FRAP}
\end{figure}
The experimental technique used to measure the slip lengths is described in details \textcolor{red}{in the supplementary materials}, and with a discussion on the resolution in \cite{henot2017}. It is an improved version of the velocimetry technique described by L\'eger {\em et al.} \cite{leger_wall_1997}. As can be seen in Figure~\ref{principe_FRAP}, the determination of the slip length relies on the observation under a simple shear of a pattern drawn in the fluorescent polymer using photobleaching. This technique allows an independent measurement the slip lengths at both surfaces and the real shear rate $\dot{\gamma}$ experienced by the polymer melt during the shear.

\begin{figure}[htbp]
  \centering
  \includegraphics[width=6cm]{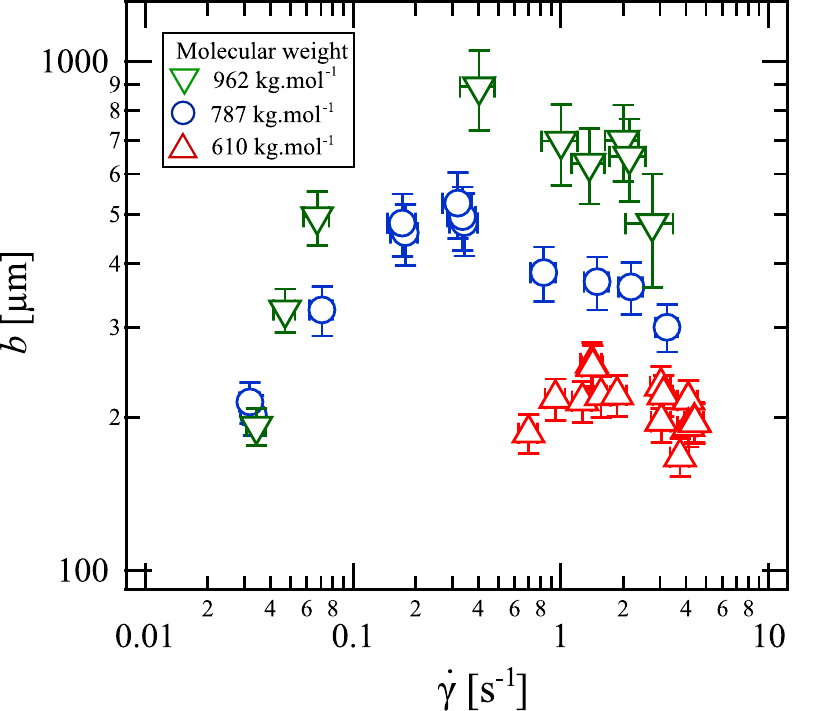}
 \caption{Slip lengths as a function of the shear rate experienced by the fluid for melts of molecular weight 609, 787 and 962~kg\(\cdot\)mol$^{-1}$.}.
  \label{glissement_brut}
\end{figure}

The slip lengths measured for the three investigated melts are reported in Figure~\ref{glissement_brut} as a function of the shear rate experienced by the fluid. Two slip regimes are observed: at low shear rates, $b$ increases with $\dot{\gamma}$ while above a critical shear rate $\dot{\gamma}^\star$ which depends on the sample, the slip lengths decrease slowly. Such a transition has yet been reported in the literature by Migler {\em et al.} \cite{migler_slip_1993} and Massey~\textit{et al.}~\cite{massey_1998} and has been attributed to few adsorbed long chains on the surface~\cite{brochard-wyart_slippage_1996} which undergo a transition from entangled to disentangled. The exact same mechanism has been recently used to explain the difference of slippage observed in dewetting compared to leveling experiments~\cite{ilton_beyond_2017}. Due to the large polydispersity of loops and tails in case of adsorbed chains, the transition appears progressive contrary to what has been observed for end grafted chains. Anyway, for large enough shear rates, the surface attached chains become fully disentangled from the melt chains leading to $b$ approximatively independent of $\dot{\gamma}$. Quite recently, Chennevi\`ere {\em et al.}~\cite{chenneviere_direct_2016} based on neutrons reflectivity experiments, could probe directly such an expulsion of the melt chains from the surface layer for high shear rates. Then, \textcolor{red}{the decreasing slip lengths for $\dot{\gamma}>\dot{\gamma}^\star$} can be considered as characteristic of a slip in the disentangled regime.

\begin{figure}[htbp]
  \centering
  \includegraphics[width=6cm]{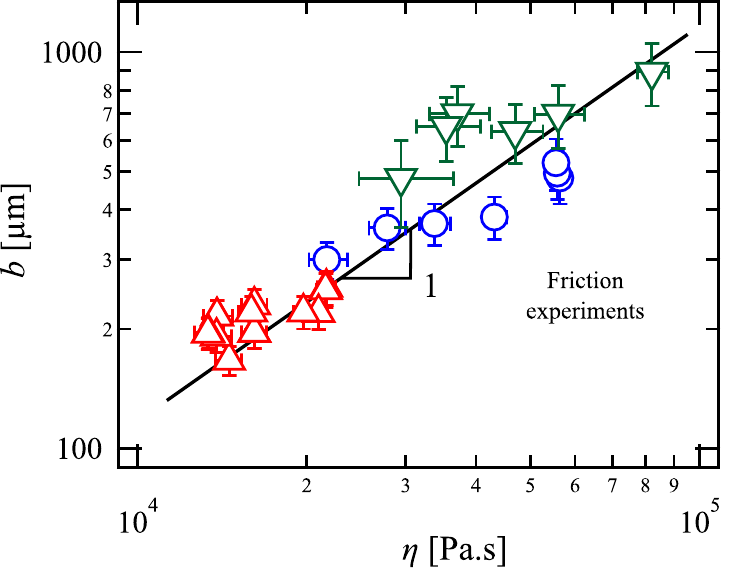}
  \caption{Slip length as a function of the viscosity of the melts at the shear rate at which $b$ was measured. The black line of slope 1 is the prediction of de Gennes's model with no adjustable parameters using the fiction coefficient obtained from friction experiments~\cite{cohen_2011}. The legend of these markers are the same than in Figure~\ref{glissement_brut}.}
  \label{bilan_glissement}
\end{figure}
Rheological measurements performed on the 787~kg\(\cdot\)mol$^{-1}$ melt showed that the shear rates in the second regime are close to the onset of shear-thinning (see supplementary materials). Similar measurements could not be done on the other melts, due to the very small available volume of polymer. The shear rate dependences of viscosity of these melts were thus estimated by extrapolation as explained in the Supplementary material. 

Figure~\ref{bilan_glissement} represents the slip lengths in the disentangled regime as a function of the viscosity at the corresponding shear rate. The vertical error bars shown on this graph come from the method of measurement. \textcolor{red}{This uncertainty on slip lengths propagates on shear rates and therefore on the determination of $\eta(\dot{\gamma})$}. The clear linear relationship between the slip length and the viscosity provides a direct proof of the full validity of the Navier's hypothesis of a linear response at the interface, and thus of a friction coefficient independent of the shear rate. It also demonstrates that the friction coefficient $k$ is indeed independent of the chain length as predicted by de Gennes in 1979~\cite{de_gennes_1979}. A linear fit of these data provides a value for the interfacial friction coefficient, with no need of any other hypothesis: $k_\mathrm{slippage}=8.6\times 10^{7}$~kg$\cdot$m$^{-1}\cdot$s$^{-1}$. It has been proposed that for such a system of a polymer flowing on a brush of the same polymer, the friction should be mainly a monomer-monomer friction \cite{ajdari_slippage_1994,brochard-wyart_slippage_1996}. We thus expect $k_\mathrm{slippage}=\zeta/a^2$, where $\zeta$ is a monomeric friction coefficient and $a$ a monomer size. Using $a=0.5$~nm, we obtain $\zeta=22\times 10^{-12}$~N$\cdot$s$\cdot$m$^{-1}$. This value is fully consistent with $\zeta=8.6\times 10^{-12}$~N$\cdot$s$\cdot$m$^{-1}$ obtained with self-diffusion measurements~\cite{leger1996} or rheological measurements~\cite{barlow_1964}.


In order to further strengthen the molecular origin of the interfacial coefficient, we have compared interfacial stress measurement on two different systems, both consisting at the molecular level PDMS monomer-monomer contact. 
\begin{figure}[htbp]
  \centering
  \includegraphics[width=6cm]{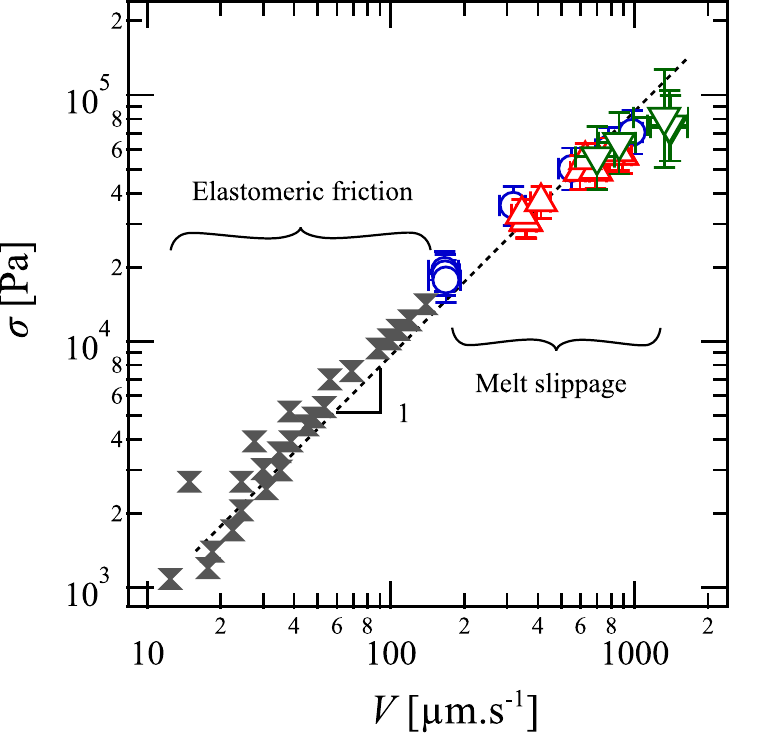}
 \caption{Tangential stress exerted by PDMS melts or crosslinked elastomers on a grafted layer of short PDMS chains. Empty symbols represent data for the melts used in this study, calculated using data of Figure~\ref{bilan_glissement}. The legend of these markers are the same than in Figure~\ref{glissement_brut}. Black symbols are data corresponding a crosslinked PDMS lens, reproduced from Cohen~\textit{et al.}~\cite{cohen_2011}.}
  \label{glissement_friction}
\end{figure}
 Bureau~\cite{bureau_2004} and Cohen~\cite{cohen_2011}, investigated the friction of small cross-linked PDMS lenses on the same grafted layers that used in this study. For soft elastomer-solid contacts, the friction is not characterized by a solid friction coefficient since the apparent and real contact areas are equal. Instead, the significant physical quantity is the tangential stress $\sigma$ exerted by the elastomer sphere on the solid surface. This stress was measured at different sliding velocities $V$ by Bureau {\em et al.} and Cohen {\em et al} \textcolor{red}{from the same group, and independently by Casoli~\textit{et al.}~\cite{casoli_friction_2001}. All the reported data are in good agreement with} a linear relationship between $\sigma$ and $V$, $\sigma(V)=\sigma_0+k_\mathrm{friction}V$, with $\sigma_0$ depending on the quality on the grafted layers. The experimental data $\sigma(V)-\sigma_0$ from~\cite{cohen_2011} are reproduced in Figure~\ref{glissement_friction} (gray markers). \textcolor{red}{At low velocity, the rubber-surface data become scattered due to their high sensibility to small heterogeneities at the surface avoiding a precise study of a possible deviation to a linear behavior.}
Using the expression $\sigma=\eta V/b$ derived from Eq.~\ref{eq_contrainte_bord} and~\ref{eq_slip_length}, the interfacial stress exerted by the three PDMS melts on the surfaces in our slip measurements were calculated, and are shown in Figure~\ref{glissement_friction} (empty colored markers). \textcolor{red}{Only the data from the disentangled regime are shown ($\dot{\gamma}>\dot{\gamma}^{*}$). All the rubber and melt friction data collapse on a single master curve. It was not possible to explore the same range of velocity with the two experiments due to contact instabilities related to the apparition of Schallamach waves at the rubber-surface interface~\cite{schallamach_how_1971}.} The dashed line in Figure~\ref{glissement_friction} is an adjustment of Eq.~\ref{eq_contrainte_bord}. 
This is the first direct proof that at the molecular level, this friction should be a monomer-monomer friction, as for both systems everything is made out of PDMS. \textcolor{red}{For these systems, we obtain $k_\mathrm{slip}=k_\mathrm{friction}\approx \zeta_0/a^2$ as suggested theoretically by~\cite{ajdari_slippage_1994,rubinstein_slippage_1993}}. 

In conclusion, using a velocimetry technique based on photobleaching, we measured the slip length for three PDMS melts of different molecular weights flowing on a weakly-adsorbing surfaces. We observed  slip lengths proportional to the melt viscosity for high enough shear rates as predicted by Navier. This linear dependence also provided a strong test of the de Gennes's model~\cite{de_gennes_1979} for polymer melts flowing on ideal surfaces at the onset of of the shear-thinning regime. More interestingly, we showed for the first time that the friction coefficient of the melts flowing on grafted layer of short PDMS chains is equal to the one measured by direct solid friction measurements of a crosslinked PDMS elastomers on the same surfaces. This allowed to directly connect for the first time solid-solid friction to solid-fluid friction in a direct measurement. 

\section*{Acknowledgements} 
 This  work  was  supported  by  ANR-ENCORE program (ANR-15-CE06-005) and European Research Council Grant (FP7/2007-2013). F.R. thank O. B\"aumchen for interesting comments on the manuscript. We thank F. Boulogne, A. Chennevi\`ere for their technical help.

\end{document}